 \documentstyle[12pt,amssymb]{article}

\newtheorem{thm}{Theorem}[section]
\def\ha{\mbox{$\cal H$}}
\def\la{\langle}
\def\ra{\rangle}

\def\ie{\emph{i.e.}}

\def\eq{\begin{equation}}
\def\en{\end{equation}}
\def\ot{\otimes}
\def\id{\mbox{\small id}}
\def\no{\mbox{\footnotesize \#}}

\def\Z{\mathbb{Z}}
\def\C{\mathbb{C}}
\def\Re{\mathbb{R}}

 \normalbaselines
\begin{document}
 \begin{center}
 \vspace*{1.0cm}

 {\LARGE{\bf Quantum Lax scheme for 
Ruijsenaars models}}

 \vskip 1.5cm

 {\large {\bf Branislav Jur\v co${}^{*}$\footnote{e-mail: jurco@mpim-bonn.mpg.de} 
and Peter Schupp${}^{**}$ }}

 \vskip 0.5 cm

${}^*$Max-Planck-Institut f\"ur Mathematik\\Vivatgasse 7\\
D-53111 Bonn, Germany\\[1ex]
${}^{**}$Sektion Physik\\
Universit\"at M\"unchen\\
Theresienstr.\ 37\\
D-80333 M\"unchen, Germany

 \end{center}

 \vspace{1 cm}

 \begin{abstract}
 We develop a quantum Lax scheme for IRF models 
and difference versions of Calogero-Moser-Sutherland
models introduced by Ruijsenaars. The construction is in the spirit of the
Adler-Kostant-Symes method generalized
to the case of face
Hopf algebras and elliptic quantum groups with dynamical
R-matrices.
 \end{abstract}

 \vspace{1 cm}

 \section{Introduction}

 The Hamiltonian of the (relativistic) Ruijsenaars model \cite{Rui}
for two particles on a line with coordinates $x_1$ and $x_2$ acts on a wave function 
in a manifestly non-local way as
\eq
\ha \,\psi(\lambda) = 
\frac{\theta(\frac{c \eta}{2} 
- \lambda)}{\theta(-\lambda)}\psi(\lambda - \eta) 
+ \frac{\theta(\frac{c \eta}{2} 
+ \lambda)}{\theta(\lambda)}\psi(\lambda + \eta), \label{Rui}
\en
where $\lambda = x_1 - x_2$ is the relative coordinate, $c \in \C$ is a coupling 
constant, $\eta$ is the
relativistic deformation parameter,  and $\theta$ is the Jacobi theta function.
The Hamiltonian apparently contains shift operators
$t_i$ that generate a one-dimensional
graph:
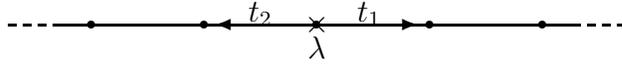
\begin{center}
\unitlength 0.50mm
\linethickness{0.4pt}
\thicklines
\begin{picture}(164.00,20.00)
\put(40.00,10.00){\circle*{2.00}}
\put(100.00,10.00){\circle*{2.00}}
\put(73.00,10.00){\vector(1,0){24.00}}
\put(67.00,10.00){\vector(-1,0){24.00}}
\put(70.00,7.00){\makebox(0,0)[ct]{$\lambda$}}
\put(84.00,11.00){\makebox(0,0)[cb]{$t_1$}}
\put(55.00,11.00){\makebox(0,0)[cb]{$t_2$}}
\put(10.00,10.00){\circle*{2.00}}
\put(130.00,10.00){\circle*{2.00}}
\put(70.00,10.00){\circle*{2.00}}
\put(70.00,10.00){\makebox(0,0)[cc]{$\times$}}
\thinlines
\put(0.00,10.00){\line(1,0){140.00}}
\put(142.00,10.00){\line(1,0){2.00}}
\put(146.00,10.00){\line(1,0){2.00}}
\put(150.00,10.00){\line(1,0){2.00}}
\put(-12.00,10.00){\line(1,0){2.00}}
\put(-8.00,10.00){\line(1,0){2.00}}
\put(-4.00,10.00){\line(1,0){2.00}}
\end{picture}
\end{center}
Relative to a fixed vertex $\lambda$ the
vertices of this (ordered) graph 
are at points $\eta\cdot\Z \in \Re$.
This picture generalizes arbitrary ordered graphs, which we 
shall consider in the following.
We will continue to
fix a vertex 
$\lambda$ to avoid a continuous
family of disconnected graphs.

\subsection{Face algebras}
 
The Hilbert space of the model 
are vector spaces on paths of fixed length on the ordered graph.
Its operators are elements of
a Face Algebra $F$ \cite{Hay1}. This is a novel mathematical structure that
naturally incorporates the complicated shifts of the Ruijsenaars model.
There are two commuting projection operators $e^i, e_i \in F$ 
onto bra's and ket's corresponding to each vertex $i$:
$
e_i e_j = \delta_{ij} e_i$, $e^i e^j = \delta_{ij} e^i$,
$\sum e_i = \sum e^i = 1$.
$F$ has a coalgebra structure such that $e^i_j \equiv e^i e_j = e_j e^i$ is a
corepresentation:
\eq
\Delta(e^i_j) = {\textstyle\sum}_k e^i_k \ot e^k_j,
\quad \epsilon(e^i_j) = \delta_{ij}; \qquad \Rightarrow \quad
\Delta(1) = \sum_k e_k \ot e^k \neq 1 \ot 1 .
\en
The latter is a key feature of face algebras and weak 
$C^*$-Hopf
algebras.
The matrix indices of $e^i_j$ are vertices,
\ie\ paths of length zero. In the given setting it is natural to also
allow paths of fixed length on a finite
oriented graph $\cal G$
as matrix indices. 
We shall use capital letters 
to label paths. A path $P$ has an origin (source) $\cdot P$,
an end (range) $P \cdot$ and a length 
$\no P$. 
Two paths $Q$, $P$ can be concatenated to form a new path
$Q \cdot P$, if the end of the first path coincides with the start of
the second path, \ie\ if 
$Q \cdot = \cdot P$.
The symbols
$T^A_B$, where $\no A = \no B = \no A' \geq 0$,
with relations
\eq
\Delta\left( T^A_B \right) = \sum_{A'} T^A_{A'} \ot T^{A'}_B, \qquad
\epsilon( T^A_B ) = \delta_{A B} \label{eps}, \qquad
T^A_B T^C_D = \delta_{A\cdot,\cdot C} \delta_{B\cdot,\cdot D} T^{A \cdot
C}_{B \cdot D} \label{tt}
\en
span an object that obeys the axioms of a face algebra.
Relations (\ref{eps}) make $T^A_B$ a
corepresentation; and the last expression is the rule for combining
representations.
The axioms of a face algebra can be found in \cite{Hay1}.
\paragraph{\it Pictorial representation:}
$$T^A_B \,\sim\;
\unitlength1mm
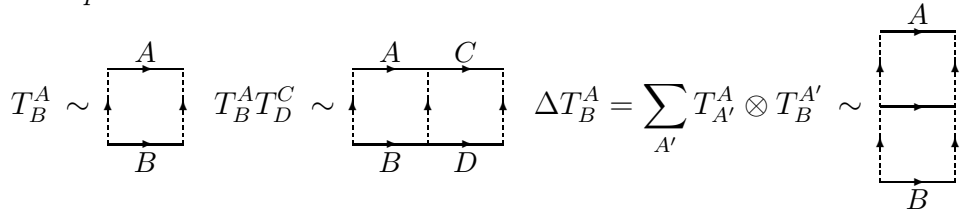
\begin{picture}(10,10)(0,4)\small
\put(0,0){\line(1,0){10}}
\put(0,10){\line(1,0){10}}
\multiput(0,0)(0,1){10}{\line(0,1){0.5}}
\multiput(10,0)(0,1){10}{\line(0,1){0.5}}
\put(0,5){\vector(0,1){1}}
\put(10,5){\vector(0,1){1}}
\put(5,0){\vector(1,0){1}}
\put(5,10){\vector(1,0){1}}
\put(5,-1){\makebox(0,0)[ct]{$B$}}
\put(5,11){\makebox(0,0)[cb]{$A$}}
\end{picture}
\quad
T^A_B T^C_D \,\sim\;
\begin{picture}(20,10)(0,4)\small
\put(0,0){\line(1,0){10}}
\put(0,10){\line(1,0){10}}
\multiput(0,0)(0,1){10}{\line(0,1){0.5}}
\multiput(10,0)(0,1){10}{\line(0,1){0.5}}
\put(0,5){\vector(0,1){1}}
\put(10,5){\vector(0,1){1}}
\put(5,0){\vector(1,0){1}}
\put(5,10){\vector(1,0){1}}
\put(5,-1){\makebox(0,0)[ct]{$B$}}
\put(5,11){\makebox(0,0)[cb]{$A$}}
\put(10,0){\line(1,0){10}}
\put(10,10){\line(1,0){10}}
\multiput(20,0)(0,1){10}{\line(0,1){0.5}}
\put(20,5){\vector(0,1){1}}
\put(15,0){\vector(1,0){1}}
\put(15,10){\vector(1,0){1}}
\put(15,-1){\makebox(0,0)[ct]{$D$}}
\put(15,11){\makebox(0,0)[cb]{$C$}}
\end{picture}\quad
\Delta T^A_B = \sum_{A'} T^A_{A'} \otimes T^{A'}_B \,\sim\;
\begin{picture}(15,10)(0,9)\small
\put(0,20){\line(1,0){10}}
\multiput(0,10)(0,1){10}{\line(0,1){0.5}}
\multiput(10,10)(0,1){10}{\line(0,1){0.5}}
\put(0,15){\vector(0,1){1}}
\put(10,15){\vector(0,1){1}}
\put(5,20){\vector(1,0){1}}
\put(5,21){\makebox(0,0)[cb]{$A$}}
\put(0,0){\line(1,0){10}}
\put(0,10){\line(1,0){10}}
\multiput(0,0)(0,1){10}{\line(0,1){0.5}}
\multiput(10,0)(0,1){10}{\line(0,1){0.5}}
\put(0,5){\vector(0,1){1}}
\put(10,5){\vector(0,1){1}}
\put(5,0){\vector(1,0){1}}
\put(5,10){\vector(1,0){1}}
\put(5,-1){\makebox(0,0)[ct]{$B$}}
\end{picture}\vspace{4mm}
$$
The dashed paths indicate the $F$-space(s). Inner paths are summed over.

It is convenient to work with
an abstract, universal $T$, which is the
canonical element $T_{12}$ of $U \ot F$, 
where $U$ is the dual of $F$ via the
pairing $\la \: , \: \ra$.\footnote{Here 
and in the following we 
shall frequently suppress the
the second index of $T$; it corresponds to the $F$-space.
The displayed expressions are
short for $\la T_{12},f\ot \id\ra = f \in F$,
$\la T_{12} T_{13}, f \ot \id \ot \id\ra 
= \Delta\, f \in F \ot F$
and $\la T_{13} T_{23}, f\ot g\ot \id\ra = f g \in F$.}
\[ 
\la T , f \ra = f, \quad \la T_1 \ot T_1, f \ra = \Delta\,f, \quad
\la T_1 T_2 , f \ot g \ra = f g ; \quad f , g \in F
\]

A face \emph{Hopf} algebra has an antipode $S$ which is denoted by a tilde
in the universal tensor
formalism: $\la \tilde T, f\ra = S(f)$. 
An ordinary Hopf algebra is a special case of a Face Hopf algebra with a single
vertex.
Ordinary matrix indices correspond
to closed loops in that case.

\subsection{Boltzmann weights}

The axioms \cite{Hay1} for a quasitriangular face algebra are similar to
those of a quasitriangular Hopf algebra; there is a universal
$R \in U \ot U$ that controls the non-cocommutativity of the coproduct in $U$
and the non-commutativity of the
product in $F$,
\eq
R T_1 T_2 = T_2 T_1 R , \quad \tilde R T_2 T_1 = T_1 T_2 \tilde R,
\quad \tilde R \equiv (S \ot \id)(R),
\label{rtt}
\en
however the antipode of $R$ is
no longer inverse of $R$ but rather 
$\tilde R R = \Delta(1)$.
The numerical 
``$R$-matrix'' obtained by contracting $R$ with two face corepresentations
is given by the face Boltzmann weight $W$:
\[ 
\la R , T^A_B \ot T^C_D \ra \; = \; R^{A C}_{B D}  
\;\equiv\; W\Big( {C {B \atop A} D} \Big)
\quad\sim\qquad
\unitlength1mm
\begin{picture}(10,7)(-2,4)\small
\put(0,0){\line(1,0){10}}
\put(0,0){\line(0,1){10}}
\put(10,10){\line(0,-1){10}}
\put(10,10){\line(-1,0){10}}
\put(0,5){\vector(0,-1){1}}
\put(10,5){\vector(0,-1){1}}
\put(5,0){\vector(1,0){1}}
\put(5,10){\vector(1,0){1}}
\put(5,-1){\makebox(0,0)[ct]{$A$}}
\put(5,11){\makebox(0,0)[cb]{$B$}}
\put(-1,5){\makebox(0,0)[rc]{$C$}}
\put(11,5){\makebox(0,0)[lc]{$D$}}
\end{picture}\]\vspace{2mm}

\noindent The Boltzmann weight is zero unless $C\cdot A$ and $B\cdot D$
are valid paths with common source and range.
For $f \in F$ there are two algebra homomorphisms
$F \rightarrow U$:
\eq
R^+(f)  = \la R, f \ot \id\ra ,\qquad R^-(f) = \la \tilde R, \id \ot f\ra . \label{rplus}
\en
$R$ satisfies the Yang\-Baxter Equation
$R_{12} R_{13} R_{23} = R_{23} R_{13} R_{12}$.
Contracted 
with $T^A_B \ot T^C_D \ot T^E_F$ this expression yields
a numerical Yang-Baxter equation with the following pictorial representation
\cite{Bax}:
\begin{center}
\unitlength 0.50mm
\linethickness{0.4pt}
\begin{picture}(145.00,46.00)
\put(0.00,20.00){\line(3,-4){15.00}}
\put(15.00,0.00){\line(1,0){25.00}}
\put(40.00,0.00){\line(3,4){15.00}}
\put(55.00,20.00){\line(-3,4){15.00}}
\put(40.00,40.00){\line(-1,0){25.00}}
\put(15.00,40.00){\line(-3,-4){15.00}}
\put(90.00,20.00){\line(3,-4){15.00}}
\put(105.00,0.00){\line(1,0){25.00}}
\put(130.00,0.00){\line(3,4){15.00}}
\put(145.00,20.00){\line(-3,4){15.00}}
\put(130.00,40.00){\line(-1,0){25.00}}
\put(105.00,40.00){\line(-3,-4){15.00}}
\put(0.00,20.00){\line(1,0){25.00}}
\put(25.00,20.00){\line(3,4){15.00}}
\put(25.00,20.00){\line(3,-4){15.00}}
\put(105.00,40.00){\line(3,-4){15.00}}
\put(120.00,20.00){\line(-3,-4){15.00}}
\put(120.00,20.00){\line(1,0){25.00}}
\put(15.00,40.00){\vector(-3,-4){9.00}}
\put(0.00,20.00){\vector(3,-4){9.00}}
\put(15.00,40.00){\vector(1,0){15.00}}
\put(0.00,20.00){\vector(1,0){15.00}}
\put(15.00,0.00){\vector(1,0){15.00}}
\put(25.00,20.00){\vector(3,-4){9.00}}
\put(40.00,40.00){\vector(-3,-4){9.00}}
\put(40.00,40.00){\vector(3,-4){9.00}}
\put(55.00,20.00){\vector(-3,-4){9.00}}
\put(90.00,20.00){\vector(3,-4){9.00}}
\put(105.00,40.00){\vector(-3,-4){9.00}}
\put(105.00,40.00){\vector(3,-4){9.00}}
\put(120.00,20.00){\vector(-3,-4){9.00}}
\put(130.00,40.00){\vector(3,-4){9.00}}
\put(145.00,20.00){\vector(-3,-4){9.00}}
\put(105.00,40.00){\vector(1,0){15.00}}
\put(120.00,20.00){\vector(1,0){15.00}}
\put(105.00,0.00){\vector(1,0){15.00}}
\put(20.00,30.00){\makebox(0,0)[cc]{$R_{13}$}}
\put(40.00,20.00){\makebox(0,0)[cc]{$R_{23}$}}
\put(20.00,10.00){\makebox(0,0)[cc]{$R_{12}$}}
\put(105.00,20.00){\makebox(0,0)[cc]{$R_{23}$}}
\put(125.00,10.00){\makebox(0,0)[cc]{$R_{13}$}}
\put(125.00,30.00){\makebox(0,0)[cc]{$R_{12}$}}
\put(72.00,20.00){\makebox(0,0)[cc]{$=$}}
\put(27.00,42.00){\makebox(0,0)[cb]{$B$}}
\put(27.00,-2.00){\makebox(0,0)[ct]{$A$}}
\put(5.00,30.00){\makebox(0,0)[rb]{$E$}}
\put(5.00,10.00){\makebox(0,0)[rt]{$C$}}
\put(50.00,10.00){\makebox(0,0)[lt]{$F$}}
\put(50.00,30.00){\makebox(0,0)[lb]{$D$}}
\put(117.00,42.00){\makebox(0,0)[cb]{$B$}}
\put(117.00,-2.00){\makebox(0,0)[ct]{$A$}}
\put(95.00,30.00){\makebox(0,0)[rb]{$E$}}
\put(95.00,10.00){\makebox(0,0)[rt]{$C$}}
\put(140.00,10.00){\makebox(0,0)[lt]{$F$}}
\put(140.00,30.00){\makebox(0,0)[lb]{$D$}}
\end{picture}
\end{center}
\vspace{1ex}
The inner edges are paths that are summed over. 

\section{Quantum factorization}

Cocommutative functions, like the trace of the T-matrix, 
provide mutually commutative operators including the Hamiltonian. 
The following theorem gives for
the case of face Hopf algebras what has become known as the
``Main theorem'' for the solution by factorization of the equations
of motion \cite{AKS,RS,SJ,PB}:

\begin{thm}[Main theorem for face algebras]
\mbox{ }

\begin{enumerate}
\item[(i)] The set of cocommutative functions, $I$, is a commutative
subalgebra of $F$.
\item[(ii)] The Heisenberg 
equations of motion defined by a Hamiltonian $\ha \in I$
are of Lax form
$i \frac{d T}{dt} = \left[ M^\pm , T \right]$,
with $M^\pm = 1 \ot \ha - m_\pm \in U_\pm \ot F$, 
$m_\pm = R^\pm(\ha_{(2)}) \ot \ha_{(1)}$; see (\ref{rplus}).
\item[(iii)] Let $g_\pm(t) \in U_\pm \ot F$
be the solutions to the factorization problem
$g_-^{-1}(t) g_+(t) = \exp( i t (m_+ - m_-)) \: \in \: U \ot F ,$
then 
$T(t) = g_\pm(t) T(0) g_\pm(t)^{-1} $
solves the Lax equation; $\:g_\pm(t)$ are given by 
$g_\pm(t) = \exp(-it(1\ot h))\,\exp(it(1\ot h - M^\pm(0))$
and are the solutions to the differential equation
$i  \frac{d}{dt}g_\pm(t) =
M^\pm(t) g_\pm(t), \;  g_\pm(0) = 1 . $
\end{enumerate}
\end{thm}

\section{Dynamical operators}

Like we
mentioned in the introduction we are interested in the action of
the Hamiltonian with respect to a fixed vertex.
We fix a vertex with the help of $e^\lambda, e_\lambda \in F$.
R-matrices with a fixed vertex are {\it Dynamical $R$-matrices}:
$
R_{12}(\lambda) \equiv  \la R , T_1(\lambda) \ot T_2 \ra,
$
where $T(\lambda)^A_B$ is zero unless  the range (end) of path $A$ is equal to the
fixed vertex $\lambda$:
\eq
T(\lambda)^A_B \; = \; T^A_B \, e^\lambda
\quad\sim\quad
\unitlength1mm
\begin{picture}(10,8)(0,4)\small
\put(0,0){\line(1,0){10}}
\put(0,10){\line(1,0){10}}
\multiput(0,0)(0,1){10}{\line(0,1){0.5}}
\multiput(10,0)(0,1){10}{\line(0,1){0.5}}
\put(0,5){\vector(0,1){1}}
\put(10,5){\vector(0,1){1}}
\put(5,0){\vector(1,0){1}}
\put(5,10){\vector(1,0){1}}
\put(10,10){\makebox(0,0)[cc]{$\times$}}
\put(11,10){\makebox(0,0)[lb]{$\lambda$}}
\put(5,-1){\makebox(0,0)[ct]{$B$}}
\put(5,11){\makebox(0,0)[cb]{$A$}}
\end{picture}
\en\vspace{2mm}

\noindent The formalism naturally includes shifts, as can e.g.\
be seen in the {\it Dynamical $RTT$-equation} \cite{Fel}
\eq
R_{12}(\lambda) T_{1}(\lambda - h_2) T_{2}(\lambda)
= T_{2}(\lambda - h_1) T_{1}(\lambda) R_{12}(\lambda - h_3) .
\en
and the {\it Dynamical Yang-Baxter equation} \cite{Ger}
\eq
R_{12}(\lambda) R_{13}(\lambda - h_2) R_{23}(\lambda)
= R_{23}(\lambda - h_1) R_{13}(\lambda) R_{12}(\lambda - h_3) .
\en
The shift operators $h_1$, $h_2$, $h_3$
can be expressed in terms of $e^\mu$, $e_\eta$ and
their duals $E^\mu$, $E_\eta$,
if we assume a (local) embedding of the vertices of the
graph in $\C^n$.

The Hamiltonian of the
Ruijsenaars model is the trace of a $T$-matrix,  in an appropriate
representation.
It can be written as a sum of operators that act in
subspaces corresponding to paths ending in a vertex
$\lambda$:
$
\ha =  \sum_\lambda \ha(\lambda)
$
with
$\ha(\lambda) = \ha e^\lambda = \sum T(\lambda)^Q_Q$.
The pictorial representation of the Hamiltonian is two closed dashed
paths ($F$-space) connected by paths $Q$ of fixed length
that are summed over. In $\ha(\lambda)$ the end of path
$Q$ is fixed. When we look at a representation on Hilbert space the
paths $Q$ with endpoint $\lambda$ that appear in the component
$\ha(\lambda)$ of the Hamiltonian $\ha$ will shift the argument
of a state $\psi(\lambda)$
corresponding to the vertex $\lambda$ to a new vertex corresponding to the
starting point of the path $Q$, exactly as in (\ref{Rui}).

Given the Boltzmann weights \cite{Has,Jim}, the Main Theorem provides Lax operators
and Lax equations for the Ruijsenaars model \cite{JS}.

 \section*{Acknowledgments}
 
We would like to thank Pavel Winternitz for
interest and support, and Koji Hasegawa and Jan Felipe
van Diejen for interesting discussions. B. J. would like to thank the Alexander von Humboldt Foundation for support.

 \end{document}